\documentclass[12pt]{article}
\usepackage{mathrsfs}
\usepackage{}
\usepackage{geometry}  
\usepackage{graphicx}
\usepackage{amssymb}
\usepackage{amsmath}
\usepackage{epstopdf}
\usepackage{comment}
\usepackage{cite}

\usepackage[hyperindex=true,
          pdfstartview=FitH,
          bookmarksnumbered=true,
          bookmarksopen=true,
          citecolor=blue,
          linkcolor=blue,
          colorlinks=true,
          unicode]{hyperref}

\allowdisplaybreaks

\begin{document}

\title{Exact black hole formation in three dimensions}
\author{Wei Xu \thanks{{\em
        email}: \href{mailto:xuweifuture@gmail.com}
        {xuweifuture@gmail.com}}\\
School of Physics, Huazhong University of Science and Technology, \\ Wuhan 430074, China}

\date{}
\maketitle

\begin{abstract}
We consider three dimensional Einstein gravity non-minimally coupled to a real scalar field with a self-interacting scalar potential and present the exact black hole formation in three dimensions. Firstly we obtain an exact time-dependent spherically symmetric solution describing the gravitational collapse to a scalar black hole at the infinite time, i.e. in the static limit. The solution can only be asymptotically AdS because of the No-Go theorem in three dimensions which is resulted from the existence of a smooth
black hole horizon. Then we analyze their geometric properties and properties of the time evolution. We also get the exact time-dependent solution in the minimal coupling model after taking a conformal transformation.
\end{abstract}

\section{Introduction}
The formation of black holes due to gravitational collapse
is the fundamental and important topic in general relativity. As it can shed light on the understanding of spacetime singularities, cosmic censorship, critical phenomenon and gravitational waves \cite{Joshi:2008zz,Gundlach:2007gc,Fryer:2011zz,Joshi:2012mk}, people always pay much attention for this study. However, it is really difficult to construct exact solutions describing the time evolution of black holes. Some results, taking the famous Vaidya metric \cite{Vaidya:1999zza,Wang:1998qx} for example, only focus on certain generically specified matter energy-momentum tensor. Recently, an improvement with great importance of this study is present in \cite{Zhang:2014sta}, which shows an exact time-dependent solution that describes gravitational collapse to a static scalar-hairy black hole in four dimensional Einstein gravity minimally coupled to a dilaton with a additional scalar potential. Then the exact black hole formation is generalized to $D=4, N=4$ Gauged Supergravity \cite{Lu:2014eta}, including the dynamic C-metrics \cite{Lu:2014ida,Lu:2014sza}.

In this paper, we focus on the exact black hole formation in three dimensions. In the absence of matter source, gravity in three dimensions is locally trivial because of the lack of propagating degrees of freedom. Beyond this relative simplicity, thank to the AdS/CFT correspondence \cite{Maldacena:1997re}, one can expect that the three dimensional dynamical AdS black holes could provide a holographic description for certain two dimensional non-equilibrium thermal systems of strongly-coupled field theories, especially for which its infinite conformal symmetries making it completely integrable. These fact make three dimensional gravity an ideal theoretical laboratory to understand the essential properties of the theory in four and higher dimensions.

We consider three dimensional Einstein gravity non-minimally coupled to a real scalar field with a self-interacting scalar potential. We obtain an exact time-dependent spherically symmetric solution in this model, describing the gravitational collapse to a scalar black hole at the infinite time, i.e. in the static limit. We also get the exact time-dependent solution in the minimal coupling model after taking a conformal transformation.
Actually, the final states, namely the three dimensional static scalar black holes are well-known and already reported in literature \cite{Henneaux:2002wm,Kolyvaris:2009pc,Correa:2010hf,Correa:2011dt} and \cite{Martinez:1996gn,Nadalini:2007qi} for minimal and non-minimal coupling model, respectively. They have
attracted considerable attentions and are generalized to charged \cite{Xu:2013nia,Mazharimousavi:2014vza,Xu:2014uka} and rotating cases \cite{Degura:1998hw,Aparicio:2012yq,Correa:2012rc,Zhao:2013isa,Xu:2014uha,Zou:2014gla} in both models. Note the time-dependent solutions can both only be asymptotically AdS because of the No-Go theorem in three dimensions which is resulted from the existence of a smooth
black hole horizon \cite{Ida:2000jh}.

The paper is organized as follows. In next section, we firstly present an exact time-dependent spherically symmetric solution and its properties in three dimensional theory that Einstein gravity non-minimally coupled to a self-interacting real scalar field. We also give the apparent horizon and the corresponding exact time-dependent solution in the minimal coupling model after taking a conformal transformation in the Appendix. In Section 3, some concluding remarks are given.

\section{The theory in non-minimal coupling model and time-dependent solution}
We consider the exact black hole formation in three dimensional theory that Einstein gravity non-minimally coupled to a self-interacting real scalar field. The action of the system reads
\begin{align}
&I=\int\mathrm{d} x^3\sqrt{-g}\left[\frac{R}{2}-\frac{1}{2}g^{\mu\nu}\nabla_{\mu}\Phi\nabla_{\nu}\Phi
-\frac{1}{16} R\Phi^2-U(\Phi)\right],\nonumber\\
&U(\Phi)=-\frac{1}{\ell^2}+\left({\frac {1}{512\ell^2}} -\frac{\alpha}{2}\right) {\Phi}^{6},
\end{align}
where $\Phi$ is the scalar field and $U(\Phi)$ is the self-coupling scalar potential. This potential is widely studied together with several static exact black hole solutions dressing by a non-minimally coupled scalar field \cite{Martinez:1996gn,Nadalini:2007qi}. Besides, the static limit of the charged case \cite{Xu:2013nia,Mazharimousavi:2014vza,Xu:2014uka,Cardenas:2014kaa} and rotating cases \cite{Zhao:2013isa,Zou:2014gla} can also reduce to this theory. There are two parameters $\ell$ and $\alpha$ in the potential. Obviously, the constant term in the scalar potential plays the role of cosmological constant $\Lambda=-\frac{1}{\ell^2}$. We begin with the negative cosmological constant because of the requirement for the existence of a smooth black hole horizon, which is actually the No-Go theorem in three dimensions \cite{Ida:2000jh}. Another parameter $\alpha$ is related to the mass of the black holes and is always positive, both for the time-independent \cite{Martinez:1996gn} and time-dependent case. Note the latter case is presented later in the paper.

Using the Eddington-Finekelstein-like coordinates, we can get the time-dependent solution, which reads
\begin{align}
\mathrm{d}s^{2}=-f(u,r)\mathrm{d}u^{2}
+2\mathrm{d}u\mathrm{d}r
+{r}^{2}\tanh \left( \,{\frac {12\alpha u}{q}} \right)^{2/3}\mathrm{d}\psi^2
\label{metric1}
\end{align}
with the metric function
\begin{align}
&f(u,r)={\frac {{r}^{2}}{{\ell}^{2}}}+\frac{8\alpha \left(\tanh \left( \,{\frac {12\alpha u\,}{q}}
\right)^{2}-1 \right) \,r}{{q}\tanh \left(
\,{\frac {12\alpha u\,}{q}}\right)}-\,{\frac {12\alpha
}{{q}^{2}}}-\frac{\alpha}{{q}^{3}{r}}\tanh \left( \,{\frac {12\alpha u\,}{q}} \right) ,
\label{fmetric}
\end{align}
where the coordinate ranges are given by $-\infty< u<+\infty$, $r\geq0$, $-\pi\leq\psi\leq+\pi$.
The scalar field behaviors as
\begin{align}
&\Phi(u,r) =\pm\sqrt {\frac{1}{qr\tanh \left( \,{\frac {12\alpha u}{q}}
\right)^{-1}+1/8}},
\end{align}
where the additional free parameter $q$, namely the ``scalar charge" characterizes the strength of the scalar field. When $q=0$, we get $\tanh \left(\,{\frac {12\alpha u\,}{q}} \right)=1$, hence $\Phi$ is the constant scalar field and the theory reduces to Einstein gravity with a slightly-corrected cosmological constant. However, the higher order curvature invariants for this case are singular as shown later. As $q$ is not a coordinate viable, one can never ``built" a horizon to surround the singularity. Hence this degenerated case is actually a naked singularity which is not physically acceptable. When $q=+\infty$, the scalar field is vanishing and the solution degenerates to the massless static BTZ vacuum solution \cite{Banados:1992wn}. On the other hand, we require $q>0$ in order to make $\Phi$ not
to be singular at finite nonzero coordinate viable $r$ when the spacetime begins to evolve (i.e. $u\geq 0$). Since the scalar field is well behaved from the apparent horizon to the asymptotic region, the physical properties should thus be maintained in the corresponding time-dependent solution in the minimal coupling model, which can be resulted from taking a conformal transformation and shown in Appendix A.

Consider the static limit $ u\rightarrow+\infty$, one can get $\tanh \left(\,{\frac {12\alpha u\,}{q}} \right)\rightarrow1$, the solution reduces to the static black hole
\begin{align}
\mathrm{d}s^{2}=-f(r)\mathrm{d}t^{2}
+\frac{\mathrm{d}r^2}{f(r)}
+r^2\mathrm{d}\psi^2
\end{align}
with the radial metric function and the scalar field
\begin{align}
&f(r)={\frac {{r}^{2}}{{\ell}^{2}}}-\,{\frac {12\alpha
}{{q}^{2}}}-\frac{\alpha}{{q}^{3}{r}},\nonumber\\
&\Phi(r) =\pm\sqrt{\frac{1}{qr+\frac{1}{8}}}.
\end{align}
Here we have taken the coordinate transformation $\mathrm{d}t=\mathrm{d}u+\frac{\mathrm{d}r}{f(r)}$ to get the above Schwarzschild-like solution. This solution is firstly studied in \cite{Martinez:1996gn}. When $q=\frac{1}{8B}$, it takes the exact same form with the static simplified case of the charged scalar black holes \cite{Xu:2013nia} and rotating scalar black holes \cite{Zhao:2013isa,Zou:2014gla}. The horizon structure is analyzed in detail in \cite{Xu:2013nia}. The existence of the event horizon and the stable system indicate the physically acceptable range for the parameter $\alpha$, i.e. $\alpha\in(0,\frac{1}{256\ell^2}]$. The mass of the static black hole is $M_0=\frac{12\alpha}{q^2}$. Hence one can obtain the mass bound of black holes as $M_0\in(0,\frac{3}{64q^2\ell^2}]$. In this bound, the exact time-dependent spherically symmetric solution  describes the gravitational collapse to a static scalar black hole at the infinite time. The thermodynamics is discussed in \cite{Martinez:1996gn} with the first law of thermodynamics holding.

Then we study the properties of the time evolution by using the luminosity distance $R=r\tanh \left( \,{\frac {12\alpha u}{q}}\right)^{1/3}$, which leads to the following form of the time-dependent solution
\begin{align}
&\mathrm{d}s^{2}=-\tanh \left( \,{\frac {12\alpha u}{q}}
\right)^{-2/3}f( u,R)\mathrm{d} u^{2}
+2\tanh\left(\,{\frac{12\alpha u}{q}}\right)^{-1/3}\mathrm{d} u\mathrm{d}R
+R^2\mathrm{d}\psi^2,\\
&\Phi( u,R) =\pm\sqrt {\frac{1}{qR\tanh \left( \,{\frac {12\alpha u}{q}}\right)^{-4/3}+1/8}},\\
&f( u,R)={\frac {{R}^{2}}{{\ell}^{2}}}-\,{\frac {12\alpha}{{q}^{2}}}\tanh \left( \,{\frac {12\alpha u}{q}}
\right)^{2/3}-\frac{\alpha }{{q}^{3}R}\tanh \left( \,{\frac {12\alpha u\,}{q}} \right)^2.\label{fmetricR}
\end{align}
From the new metric function $f(u,R)$ in this case, one can get the effective time-dependent ``Vaidya mass" measured at infinity as
\begin{align}
M(u)=\frac{12\,\alpha}{{q}^{2}}\tanh \left(\,{\frac { 12\,\alpha u}{q}} \right)^{2/3},\label{mass}
\end{align}
which is proportional to the parameter $\alpha$. Hence the metric (\ref{metric1}) with vanishing $M( u)$ (i.e. $\alpha$) is the vacuum solution in the Einstein spacetime as the scalar field is vanishing.

When $ u\rightarrow+\infty$, we get $M( u)\rightarrow{M}_{0}=\frac{12\alpha}{q^2}$. This makes us able to estimate the characteristic relaxation time approaching the static limit, although it actually takes infinite $u$-time to reach the static black hole. Denote $y=e^{\frac{12\alpha u}{q}}$, then it leads to that $\tanh\left(\frac{12\alpha u}{q}\right)=\frac{y^2-1}{y^2+1}$, and one can get
\begin{align}
\frac{M( u)}{M_0}=\left(\frac{y^2-1}{y^2+1}\right)^{2
/3}=1-\frac{4}{3\,{y}^{2}}+O \left( \frac{1}{{y}^{4}} \right)  \simeq 1-\frac{4}{3}\,{{e}^{-\,{\frac {24\alpha u\,}{q}}}}
\end{align}
at large $u$. Then we obtain
\begin{align}
  M( u)\simeq\left(1-\frac{4}{3}\,{{e}^{-\,{\frac {2 u\,}{ u_0}}}}\right)M_0,
\end{align}
where we define the characteristic relaxation time $ u_0=\frac{q}{12\alpha}=\frac{1}{6}\sqrt{\frac{3}{\alpha\,M_0}}$ such that the spacetime reaches equilibrium at the rate of $\,{{e}^{- u/ u_0}}$.  Especially note that the bigger static mass $M_0$ always leads to the shorter relaxation time $ u_0$. This also happen in the study of black hole formation in four dimensions \cite{Zhang:2014sta}, which is generally not seen in the Vaidya spacetime.

Here we also calculate some geometric quantities to understand the geometric characteristics of
the time-dependent solution. Firstly we find that some of the components of the Cotton tensor, e.g.
\begin{align}
C_{R u R}=-\,{\frac {3\alpha}{2{q}^{3}{R}^{4}}}\tanh \left( \,{\frac {12\alpha u}{q}}\right)^{5/3}
\end{align}
are nonvanishing if $\alpha\neq0$, signifying that the metric is non conformally flat \cite{Garcia}. On the other hand, though the Ricci scalar is regular and takes the form as $R_{\mu}{}^{\mu}=-\frac{6}{\,{\ell}^{2}}$ indicating the AdS spacetime,
other higher order curvature invariants such as
\begin{align}
&R_{\mu\nu}R^{\mu\nu}=\frac{12}{\,{\ell}^{4}}+\,{\frac {3{\alpha}^{2}}{2{q}^{6}{R}^{6}}}\tanh \left( \,{\frac {12\alpha u}{q}}\right)^{4}\\
&R_{\mu\nu\lambda\sigma}
R^{\mu\nu\lambda\sigma}=\frac{12}{\,{\ell}^{4}}+\,{\frac {6{\alpha}^{2}}{{q}^{6}{R}^{6}}}\tanh \left( \,{\frac {12\alpha u}{q}}\right)^{4}
\end{align}
have an essential singularity at $R=0$ whenever $\alpha\neq0$. Since $\alpha>0$ (i.e. $M(u)>0$), the metric function $f( u,R)$ comes across the radial  horizontal axis once at least, for the reason that $f( u,+\infty)\rightarrow+\infty$, $f( u,0)\rightarrow-\infty$.
Thus, the solution (\ref{metric1}) with positive ``Vaidya mass" always contains a apparent horizon. Actually, to get the apparent horizon in detailed, one can solve analytically the roots of metric function $f( u,R)$ (Eq.\ref{fmetricR}), which is the cubic polynomial equations. As the form of the apparent horizon is very complicated, we only present it in Appendix B. Besides, the case with $q=0$ is also an essential singularity, other than the solution of theory that Einstein gravity coupled to constant scalar field.

Finally, we introduce the solution in another coordinate system, in which we use the scalar field $\Phi$ as a coordinate to replace $r$. The new form reads as
\begin{align}
\mathrm{d}s^{2}=-f( u,\Phi)\mathrm{d} u^{2}
-\frac{4\,}{ {q}{\Phi}^{3}}\tanh \left( \,{\frac {12\alpha u}{q}} \right)\mathrm{d} u\mathrm{d}\Phi
+{\frac {\, \left({\Phi}^{2} -8\right) ^{2}}{64{q}^{2}{
\Phi}^{4}}}\tanh\left(\,{\frac { 12\alpha u}{q}} \right) ^{8/3}\mathrm{d}\psi^2
\label{metric2}
\end{align}
with the metric function
\begin{align}
f(u,\Phi) &={\frac {(1-256\alpha\,{\ell}^{2})(\,{\Phi}^{2}-24)\,{\Phi}^{4}+192\,{\Phi}^{2}-512}{64{q}^
{2}{\ell}^{2} {\Phi}^{4}\left( {\Phi}^{2}-8 \right) }}\nonumber\\
&+{\frac{\,\left({\Phi}^{2}-8\right)}{64{\Phi}^{4}{q}^{2}{\ell}^{2}}} \bigg( (256\alpha{\ell}^{2}-1){\Phi}^{2}+8 \bigg)  \left( 1-\tanh \left( \,{\frac {12\,\alpha u}{q}} \right)^{2} \right).
\end{align}
Here the infinity is located at $\Phi=0$. For this case, one can find that the scalar field $\Phi$ does not evolve barely with the ``time" $u$, which may make it useful in the future study.

Note the solution with $u=0$ is not included in the above analysis. Consider the time evolution of the scalar field, it is vanishing at the initial time $u=0$ of spacetime, then is gradually condensed and finally evolves into that of the static scalar black hole at $ u=+\infty$. For the metric function of this case, it is regular and actually is that of the three dimensional AdS vacuum, which is also discussed in Appendix B. Besides, when $u=0$, the regular Ricci scalar and higher order curvature invariants indicate that the spacetime is really AdS and there is no essential singularity. Thus the birth of the evolving solution in the non-minimal coupling model can be understood as a vacuum.

\section{Conclusion}
In this paper, we consider three dimensional Einstein gravity non-minimally coupled to a real scalar field with a self-interacting scalar potential and point out the exact black hole formation in three dimensions. We have obtained an exact time-dependent spherically symmetric solution describing the gravitational collapse to a scalar black hole in the static limit. Then we analyze their geometric properties and properties of time evolution. Contrasting to the case in the well-known Vaidya spacetime, the bigger static mass $M_0$ always leads to the shorter relaxation time $ u_0$. The solution has been introduced into the form with barely un-evolving scalar field. The birth of the evolving solution can be understood as a vacuum. We also get the exact time-dependent solution in the minimal coupling model after taking a conformal transformation.

Three dimensional exact time-dependent solutions can only be asymptotically AdS because of the No-Go theorem in three dimensions which is resulted from the existence of a smooth black hole horizon. Hence the AdS/CFT correspondence will lead people to expect that the three dimensional dynamical AdS black holes could provide a holographic description for certain two dimensional non-equilibrium thermal systems of strongly-coupled field theories. For example, it is interesting to consider the holographic superconductor in the time-dependent background spacetime. These are left as future tasks.

\section*{Acknowledgements}
Wei Xu would like to thank professor Jian-wei Mei and Hong L\"u for useful conversations. This work was supported by the
Research Innovation Fund of Huazhong University of Science and Technology (2014TS125).

\section*{Appendix A: the theory in minimal coupling model and time-dependent solution}
Based on the time-dependent solution (\ref{metric1}) in non-minimal coupling model, we can obtain the corresponding solution in minimal coupling model through the following conformal transformation
\begin{align}
& u(\Phi)= V(\phi)\Omega^{-3},\quad g_{\mu \nu }=\Omega ^{2}\hat{g}_{\mu \nu },\nonumber\\
&\Omega =\cosh^2(\frac{\phi}{\sqrt{8}}),\quad\quad \Phi=\sqrt{8}\tanh(\frac{\phi}{\sqrt{8}}).
\label{trans}
\end{align}
where $\phi$, $V(\phi)$ and $\hat{g}_{\mu \nu }$ correspond to the scalar field, scalar potential and components of metric in the minimal coupling theory, respectively. $\Omega$ is the conformal transformation. Then we turn to the minimally coupling model with the action
\begin{align}
&\hat{I}=\int \mathrm{d}^{3}x\sqrt{-\hat{g}}\left[
\frac{\hat{R}}{2}- \frac{1}{2}\nabla_{\mu} \phi\nabla^{\mu} \phi -V(\phi )\right],\nonumber\\
&V(\phi)=\left(\frac{1}{\ell^2}-256\,\alpha\right)\sinh \left(\frac{\phi}{2\sqrt{2}}\right)^{6}-\frac{1}{\ell^2}\cosh \left(\frac{\phi}{2\sqrt{2}}\right)^{6}.
\end{align}
The scalar potential $V(\phi)$ has been studied in \cite{Henneaux:2002wm,Kolyvaris:2009pc,Correa:2010hf,Correa:2011dt}, especially for the static black hole solutions of the theory. Based on these static solutions, it is also generalized to the charged case \cite{Cardenas:2014kaa} and rotating cases \cite{Correa:2012rc,Xu:2014uha,Zou:2014gla}. Similarly, $\alpha$ is related to the mass of the time-independent  black holes \cite{Cardenas:2014kaa,Xu:2014uha,Zou:2014gla}, as well as the time-dependent case which is shown later. Hence we choose $\alpha\geq0$. After making a Taylor series
expansion for the scalar potential, one can find that the zeroth order term, i.e. $V(0)$ plays the role of cosmological constant $\Lambda=-\frac{1}{\ell^2}$, which is again required by the No-Go theorem in three dimensions \cite{Ida:2000jh}. The time-dependent solution in minimal coupling model reads
\begin{align}
\mathrm{d}s^{2}=-\frac{64{q}^{2}H\left( u,r \right)^{2}f( u,r)\mathrm{d} u^{2}}{ \left(8q \,H \left(  u,r \right)+\tanh\left(\frac{12\alpha u}{q}\right)\right) ^{2}}
+\frac{8qH( u,r)\mathrm{d} u\mathrm{d}r}{4qH( u,r)+\tanh\left(\frac{12\alpha u}{q}\right)}
+r^2\tanh\left(\frac{12\alpha u}{q}\right)^{2/3}\mathrm{d}\psi^2
\end{align}
with the scalar field and metric function
\begin{align}
\phi( u,r)&=2\sqrt{2}{\rm arctanh}\left(\sqrt{\frac{1}{1+8qH( u,r)\tanh\left(\frac{12\alpha u}{q}\right)^{-1}}}\right),\\
f( u,r)&={\frac
{H \left(  u,r \right)^{2}}{{\ell}^{2}}}+\,
\,{\frac {32\alpha \left( {\tanh\left(\frac{12\alpha u}{q}\right)}^{2}-1 \right) H \left(  u,r \right) }
{q\tanh\left(\frac{12\alpha u}{q}\right)}}+\,{\frac {3\alpha}{{q}^{2}}}\left({\tanh\left(\frac{12\alpha u}{q}\right)}^{2}-5\right)\nonumber\\
&-\,{\frac {3\alpha\tanh\left(\frac{12\alpha u}{q}\right) \left( {\tanh\left(\frac{12\alpha u}{q}\right)}^{2}-1
\right) }{{q}^{2} \left( 4q\,H \left(  u,r \right) +\tanh\left(\frac{12\alpha u}{q}\right) \right) }}-{
\frac {\tanh\left(\frac{12\alpha u}{q}\right)}{{q}^{3}H \left(  u,r \right) }},
\end{align}
where
\begin{align}
H( u,r)=\frac{1}{2}\left(r+\frac{\sqrt{2}}{2\sqrt{q}}\sqrt{r\left(\tanh\left(\frac{12\alpha u}{q}\right)+2kr\right)}\right).\label{phisol}
\end{align}
The scalar ``charge" behaviors similarly as it is in the non-minimal coupling model, thus we will only consider the case with $q>0$.

When $ u\rightarrow+\infty$, $\tanh\left(\frac{12\alpha u}{q}\right)\rightarrow1$, the solution reduces to the static black hole
\begin{align}
&\mathrm{d}s^{2}=-\frac{64{q}^{2}H(r)^{2}}{ \left(1+
8q\,H(r)  \right) ^{2}}f(r)\mathrm{d} u^{2}
+\frac{8qH(r)}{4qH(r)+1}\mathrm{d} u\mathrm{d}r
+r^2\mathrm{d}\psi^2,\nonumber\\
&\phi(r) =2\sqrt{2}{\rm arctanh}\left(\sqrt{\frac{1}{1+8qH(r)}}\right),\nonumber\\
&f(r)={\frac
{H(r)^{2}}{{\ell}^{2}}}-\alpha\,
 \left(\,{\frac {12}{{q}^{2}}}+{
\frac {1}{{q}^{3}H(r)}} \right),
\end{align}
where
\begin{align}
  H(r)=\frac{1}{2}\left(r+\frac{\sqrt{2}}{2\sqrt{q}}\sqrt{r(1+2qr)}\right).
\end{align}
After taking the coordinate transformation $\mathrm{d}t=\mathrm{d}u+\frac{\mathrm{d}r}{f(r)}$ which leads to the Schwarzschild-like solution, this solution is the one firstly studied in \cite{Henneaux:2002wm}. If one choose $q=\frac{1}{8B}$, it reaches exactly to the static degenerated case of rotating cases \cite{Xu:2014uha,Zou:2014gla}. The horizon structure is analyzed in detail in \cite{Xu:2014uha}, as well as its thermodynamics and the first law of thermodynamics, in which the mass of the static black hole is shown as $M_0=\frac{12\alpha}{q^2}$. Namely, the final state of the time-dependent solution is the static scalar black holes.

Similarly, using the luminosity distance $R=r\tanh \left( \,{\frac {12\alpha u}{q}}\right)^{1/3}$ to rewrite the time-dependent solution and consider the asymptotical behavior at infinity, we can get
the effective time-dependent ``Vaidya mass" as
\begin{align}
M(u)={\frac {\alpha}{{q}^{2}}}\left(\,{\tanh\left(\frac{12\alpha u}{q}\right)}^{8/3}+11\,{\tanh\left(\frac{12\alpha u}{q}\right)}^{2/3}\right).
\end{align}
When $ u\rightarrow+\infty$, we get $M\rightarrow{M}_{0}=\frac{12\alpha}{q^2}$, which is the mass of the static black hole. Note this ``Vaidya mass" seems to be different from that (\ref{mass}) in the non-minimal coupling model, which is actually owing to the scalar field contribution. Since the scalar field in both models are well behaved from the apparent horizon to the asymptotic region, the physical properties of the Einstein frame and the Jordan frame should always be equivalent.

\section*{Appendix B: The apparent horizon of time-dependent solution in non-minimal coupling model}
In order to get the apparent horizon $R=R_0>0$ of time-dependent solution in non-minimal coupling model,  defined by $0=g^{\mu\nu}\partial_{\mu}R\partial_{\nu}R$, we focus on the roots of the metric function
Eq.(\ref{fmetricR}), which can be simplified as
the following cubic polynomial equations
\begin{align}
R^3-3m^2R+2n=0,\quad\,m>0,
\end{align}
where
\begin{align*}
m=\,{\frac{2\sqrt{\alpha}{\ell}}{{q}}}\,{\tanh\left(\frac{12\alpha u}{q}\right)}^{1/3},
\quad\,n=-{\frac{\alpha{\ell}^{2}}{2{q}^{3}}}\,{\tanh\left(\frac{12\alpha u}{q}\right)}^{2}.
\end{align*}
Then we can obtain the exact roots as
\begin{align}
R_{i}=2m\sin\left(\frac{1}{3}\arcsin\left(\frac{n}{m^3}\right)+\frac{2\xi_{i}\pi}{3}\right), \quad\,i=1,2,3,\label{horizon}
\end{align}
where $\xi_{i}=0,\pm1$ respectively. When $|n|\leq\,m^3$, three roots are real and the biggest one is the apparent horizon. When $|n|>\,m^3$, we can get only one positive root representing the apparent horizon.

When $ u\rightarrow+\infty$, the apparent horizon will approach the event horizon. Actually, the form of horizons reduce to
\begin{align}
r_{i}=2m\sin\left(\frac{1}{3}\arcsin\left(\frac{n}{m^3}\right)+\frac{2\xi_{i}\pi}{3}\right),
\end{align}
where the parameters are
\begin{align*}
m=\,{\frac{2\sqrt{\alpha}{\ell}}{{q}}},
\quad\,n=-{\frac{\alpha{\ell}^{2}}{2{q}^{3}}}.
\end{align*}
This is the exact black hole horizons of the three dimensional static scalar black hole and the biggest one is the event horizon \cite{Martinez:1996gn,Nadalini:2007qi}.

Consider the birth of the spacetime $ u\rightarrow 0$, one can find that the metric function Eq.(\ref{fmetricR}) is regular and actually is that of the three dimensional AdS vacuum with the horizon Eq.(\ref{horizon}) reducing to $R=0$.

\providecommand{\href}[2]{#2}\begingroup
\footnotesize\itemsep=0pt
\providecommand{\eprint}[2][]{\href{http://arxiv.org/abs/#2}{arXiv:#2}}
\endgroup

\end{document}